# Determining Ionizing Doses in Medium Earth Orbits Using Long-Term GPS Particle Measurements

Yue Chen
Los Alamos National Laboratory
P.O. Box 1663, D436
Los Alamos, NM 87544
cheny@lanl.gov

Matthew R. Carver
Booz Allen Hamilton
8283 Greensboro Dr.
McLean, VA 22102
carver_matthew@bah.com

Steven K. Morley
Los Alamos National Laboratory
P.O. Box 1663, D436
Los Alamos, NM 87544
smorley@lanl.gov

Andrew S. Hoover
Los Alamos National Laboratory
P.O. Box 1663, B241
Los Alamos, NM 87544
ahoover@lanl.gov

*Abstract*— We use long-term electron and proton in-situ measurements made by the CXD particle instruments, developed by Los Alamos National Laboratory and carried on board GPS satellites, to determine total ionizing dose (TID) values and daily/yearly dose rate (DR) values in medium Earth orbits (MEOs) caused by the natural space radiation environment. Here measurement-based TID and DR values on a simplified sample geometry-- a small (with a radius of 0.1 mm) Silicon detector within an Aluminum shielding sphere with a thickness of 100 mil—are compared to those calculated from empirical radiation models. Results over the solar cycle 24 show that electron TID from measurements in GPS orbit is well above the values calculated from the median/mean fluences from AE8 and AE9 models, but close to model fluences at high percentiles. Also, it is confirmed that in MEOs proton contributions to TID are minor and mainly dominated by solar energetic protons. Several factors affecting those dose calculations are discussed and evaluated. Results from this study provide us another out-of-sample test on the reliability of existing empirical space radiation models, and also help estimate the margin factors on calculated dose values in MEOs that pass through the heart of the Earth's outer radiation belt.

## TABLE OF CONTENTS

## 1. INTRODUCTION

Man-made satellites operating in space are continuously exposed to harsh radiation environment. For those orbiting in the Earth's magnetosphere, energetic electrons and protons trapped inside of the Van Allen radiation belts as well as solar origin particles from flares and coronal mass ejections are the main sources for ionizing radiation. Among space radiation effects, total ionizing dose (TID) accumulated over time is well-known to cause degradation of satellite materials and electronic devices on board [1]. In addition, extreme dose rates (DR) due to severe space weather events, e.g., high fluxes of solar energetic particles impinging the Earth and recurrent relativistic (MeV) electron events in the outer radiation belt, also lead to malfunctions of semiconducting devices, including such as threshold voltage shifts, increased leakage current and power consumption, and ultimately circuit failures [2].

Therefore, reliably determining ionization dose values is essential for satellites' mission design and normal operation. Currently, ionizing effects are often quantified by running radiation transport Monte-Carlo codes with inputs of radiation particle fluences, which are usually anticipated by empirical and statistical space radiation models with a given satellite orbit as well as the mission duration. Testing with different shielding configurations and accordingly selecting proper radiation hardened devices, one can mitigate the adverse effects of ionizing radiations to a tolerable level with an optimized design, and meanwhile still meet the mission expectation with a balance of weight/size, functionality and budget. However, there are two potential issues with the above standard procedure. First, uncertainties in particle fluences from empirical models will propagate through and the final calculated (also expected) TID values will deviate from "real" ones. Second, how well extreme DRs over different time scales (e.g., daily and yearly) can be predicted is of great interest to satellite/instrument designers and operators.

One solution to the above issues is to use high-quality, continuous, long-term in-situ particle measurements. Previously in [3], TIDs in geosynchronous (GEO) orbit have been successfully calculated from 11-year-long electron and proton measurements made by Los Alamos National Laboratory (LANL) GEO satellites. Indeed, many interesting features, including how daily dose rates (dDRs) of electrons vary dramatically during space weather events and yearly dose rates (yDRs) fluctuate with a factor ~10 in solar cycle 24, have been revealed with unprecedented details. It is also shown that the average electron yDR in GEO is only about a half or quarter of the values calculated from AE8 [4] and AE9 [5] mean distributions. The work [3] focuses on the GEO that locates at the outer rim of the radiation belts with relatively weak geomagnetic field, and it would be natural to wonder what it will be like for other popular orbits.

Here we expand our ionizing dose study to the medium Earth orbits (MEOs) by using the in-situ particle measurements from another set of LANL space environment instruments on board the Global Positioning System (GPS) satellite constellation, and measurement-based TID values are again compared to those calculated from empirical trapped radiation models, including AE8, AP8 [6], AE9 and AP9 [5] as well as the statistical energetic solar particle model Emission of Solar Protons (ESP) model [7, 8]. MEO region, the space loosely defined between the Low-Earth-Orbits (LEO, with altitudes ~ 300 -1500 km) and GEO (with the altitude ~36,000 km), is an ideal place to host global wireless and navigation satellite systems. Among them the GPS constellation is one key space infrastructure that serves our modern society. However, the MEO region has remained largely underexplored due to the tendency of placing most satellites in LEO and GEO so as to avoid the most intense area of the radiation belts.

## 2. GPS NS59 Particle Fluxes, Geometry, and GRAS Model

GPS orbit—a typical circular MEO with an altitude ~20,200 km and nominal inclination of 55°—crosses the center of electron outer radiation belt four times daily, and thus experiences a very hostile radioactive environment that accounts for many anomalous radiation effects compared to other orbits. The GPS constellation has a long history of carrying energetic particle instruments developed by LANL since the 1980s [9]. Measurements from these instruments have been widely examined in a range of scientific research on the natural space radiation environment, e.g., [10, 11, 12, 13, and 14], but this is the first time the measurements being directly used for dose calculation. Recently, LANL has released decades of GPS particle data to the public [15] as part of national efforts of enhancing space-weather preparedness, and they (including those used in this work) can be retrieved from https://www.ngdc.noaa.gov/stp/space-weather/satellite-data/satellite-systems/gps/.

Particle fluxes used here are from the latest generation of charged particle instruments called Combined X-ray Dosimeter (CXD, [16]) developed by LANL. This instrument measures both electrons (with energies from 120 keV to more than 5 MeV) and protons (from ~6 MeV up to greater than 75 MeV). Readers are referred to [16] for detailed instrument description. This study examines the data from GPS satellite ns59, and its daily averaged electron and proton flux distributions over two years (2016-2017) are presented in Figure 1 as one example. Due to the inclined orbit, the measured fluxes depend on both energy and L-shell. Electron (proton) fluxes at L = 4.6 (7.2) in Panel A (D) are sorted as a function of the nominal energy range of CXD instrument. Increments of electrons in MeV electron events can be seen in Panel A closely related to geomagnetic storms indicated by the dips of the Dst index plotted in Panel C, while high fluxes for > ~ 10s MeV protons from solar energetic proton (SEP) events are also visible in Panel D (e.g., the major one on day 621). Radial profiles of 6 MeV electrons are shown in Panel B, and multiple MeV electron events can be identified at L-shells between ~ 4 - 6. Interestingly, similar enhancements are also seen for 10 MeV protons as in Panel E, although most of the time the proton channel is measuring the background. Considering the temporal and spatial resemblance between electron and proton enhancement features in Panels B and E, it is very likely this proton channel (and others) are contaminated by high fluxes of penetrating MeV electrons.

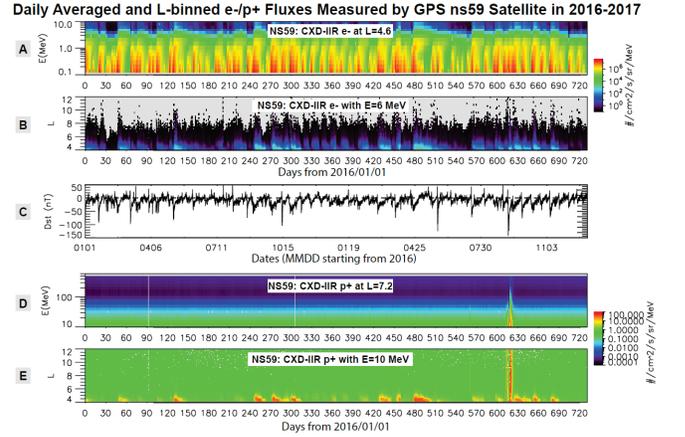

Figure 1. Overview of electron and proton dynamics observed by GPS ns59 in 2016 and 2017. Panels A and B (D and E) are electron (proton) fluxes sorted as a function of energy and L-shell, and Panel C shows the Dst index. Data gaps exist for both electrons and protons (e.g., on day 91).

For ionization dose calculations, fluences can be calculated by integrating the fluxes, as shown in Figure 1, first over the space with the consideration of how long time the satellite samples each L-shell, and then over the given time interval. In average, ns59 satellite spends 76.7% (81.9%) of daily time at L-shells ≤ 12 (15). Also note that GPS satellites cannot access L-shells below ~ 4, which is determined by their altitude at equatorial crossings. Examples of L-shell averaged electron and proton fluxes measured by ns59 in each day during 2016-2017 are presented in Figure 2. Here

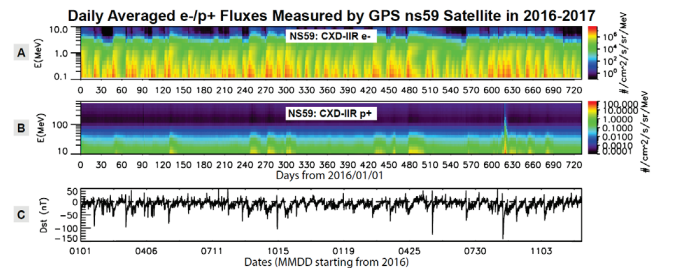

Figure 2. Overview of L-shell and daily averaged electron and proton fluxes observed by GPS ns59 in 2016 and 2017. Panel A (B) are electron (proton) daily fluxes averaged over L-shells, and Panel C shows the Dst index.

Panel A exhibits electron dynamics that are clearly associated with the geomagnetic index, resembling Figure 1A. For protons in Panel B, both the effects of MeV electrons and solar protons can be seen, while the former is distinguishable by being mostly confined to 10s MeV protons (e.g., the three events between days 240-330) while the latter is spike-like and significant up to 100s MeV. Daily energy-differential fluences are obtained by multiplying the fluxes in Figure 2 with the time factor of 86400 and angular factor of $4\pi$, and the sum of all daily fluences yields the total fluence over two years.

Like our previously work [3], a three-dimensional solid sphere geometry is adopted for this study. Our goal is to calculate the DRs and accumulated TIDs deposited on a small (with a radius of 0.1 mm) Silicon (Si) detector inside an Aluminum (Al) shielding sphere with a thickness of 100 mil. This symmetric geometry is much simplified but still suits for this demonstration study. The shielding thickness of 100 mil is selected so that results can be directly compared to those at GEO [3].

The Geant4 Radiation Analysis for Space (GRAS) model [17] is the numerical model used here to calculate ionization doses, same as in [3]. This Geant4 based, modular radiation transport code was developed for space environment effects simulation, and its capability of joint Monte Carlo tracing makes it a fast and versatile tool for radiation effects.

### 3. CALCULATE TOTAL IONIZING DOSES AND DOSE RATES

In this section, using the two-year in-situ electron and proton measurements from GPS ns59, as in Figure 2, as well as statistical space radiation specification models, we describe step by step how TID and dose rates are calculated.

In Figure 3A, the measured electron fluence spectrum is plotted in black, along with four other spectra from electron specification models AE8 and AE9: purple calculated from AE8 model median fluences (AE8-50%), aqua from AE8 with 97.7% confidence level (AE8-97.7%), green from AE9 mean values, and red from AE9 fluxes at 95% percentile. Specification model fluxes are derived by providing GPS ns59 orbital parameters and temporal information to SPENVIS [18], in which confidence levels have been added to AE8 models by considering standard deviations of fluxes and AE9 has built-in flux distributions for different percentiles (hereinafter we sometimes treat median, mean and average interchangeable for the simplicity of description). Note the measured electron spectrum at energies above 1 MeV has higher fluences than those from model average values but close to the 95th percentiles from AE9. The fluence spectra in Figure 3A are used as inputs to GRAS for dose calculations.

The same superposition principle as in [3] is applied to speed up dose calculations for this study. First, we determined the electron dose contribution function as shown in Figure 3B for the specific geometry. Each data point in this curve is obtained by feeding mono-energetic electrons with a unit fluence level of 1 cm$^{-2}$ to GRAS to calculate the corresponding ionization dose. The derived curve appears a reversed Bragg curve: low-energy electrons barely contribute to ionization, and a small peak is seen at ~ 2 MeV, and then a nearly flat distribution extending up to 10 MeV. This curve can be understood from the shape of electron's linear energy transfer (LET) curve inside Si: electrons cannot penetrate through the 100 mil Al shielding until reach at least ~ 1 MeV, and then electrons with residual energies of ~ 100s keV have high LET which explains the peak, and then LET at even higher energy does not vary much which explains the plateau-like distribution.

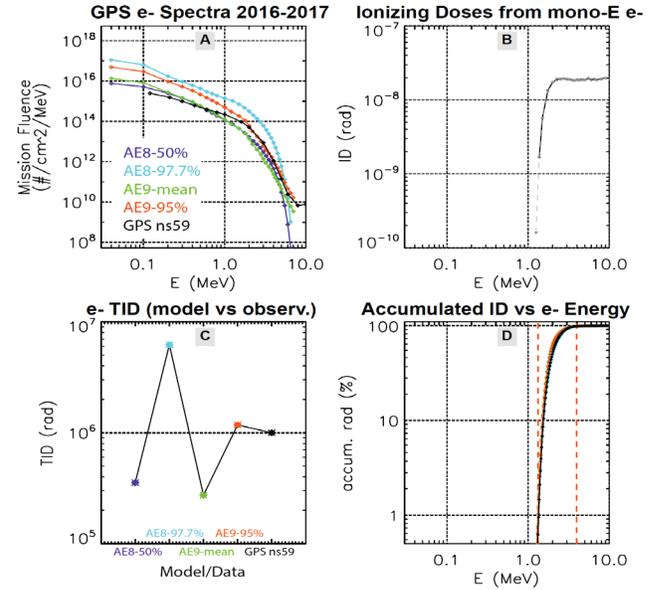

Figure 3. Calculate electron TID over 2016-2017. **A)** Electron fluence spectra, **B)** dose contribution function curve, **C)** TID calculated from different electron fluence inputs, and **D)** effective energy range determined from normalized accumulative dose percentage curves.

Using the derived dose contribution function, one can quickly calculate electron ionization dose by integrating the product of this function and any given electron fluence spectrum over the whole energy range. The advantage of this superposition method is that it greatly shortens the calculation time by more than one order of magnitude when compared to starting a full-scale GRAS run each time. Precision of this method has been tested in [3] and deemed high enough for this study.

Two-year electron TID values for ns59 calculated in this way are shown in Figure 3C. The electron TID from measured fluence is 0.999 Mrad which is 2.8 (3.7) times of the 0.355 Mrad (0.273 Mrad) calculated from AE8 (AE9) median (mean) fluences, while the TID from AE8 with a confidence level of 97.7% (AE9 with percentile of 95%) is even 6.21 Mrad (1.18 Mrad) that is ~ 16 (~ 3.3) times higher than that from the average. These results also suggest that AE8 median fluence of energetic electrons is slightly higher than that of AE9 mean in GPS orbit, and the statistical range of AE8 is wider, too.

It is also useful to determine the effective energy range for electrons in GPS orbit for this specific geometry. As in Figure 3D, normalized accumulative dose curves are plotted as a function of electron energy for each given electron spectrum. If defining the effective energy range as the core population of electrons contributing to TID from 1% to 99%, this range is ~ 1.3 – 4 MeV for electrons with a typical spectrum shape in GPS orbit as seen in Figure 3D. Electron fluences outside this range are basically ignorable for TID.

Continuous measurements by ns59 allow us to further determine daily dose rates (dDR) and track how ionization dose being accumulated over time. Figure 4A plots the dose accumulation curve based on measurements in black, in which abrupt increments (two examples marked out with arrows) are seen to be caused by high dDRs (the two corresponding spikes) in Panel B associated with MeV electron events (e.g., the vertical green strips) as in Panel C. In comparison, doses calculated from specification models accumulate with constant rates. Electron TID values from different inputs are presented in the right side of Panel A. The average electron dDR from measurements is 1.37 krad/day, compared to the mean dDR of 0.485 (0.374) krad/day for AE8-50% (AE9-mean). The highest dDR from measurements is ~6.7 krad/day, which is still within the range of 8.50 krad/day dDR set by AE8-97.7% but already beyond the dDR of 1.61 krad/day set by AE9-95%.

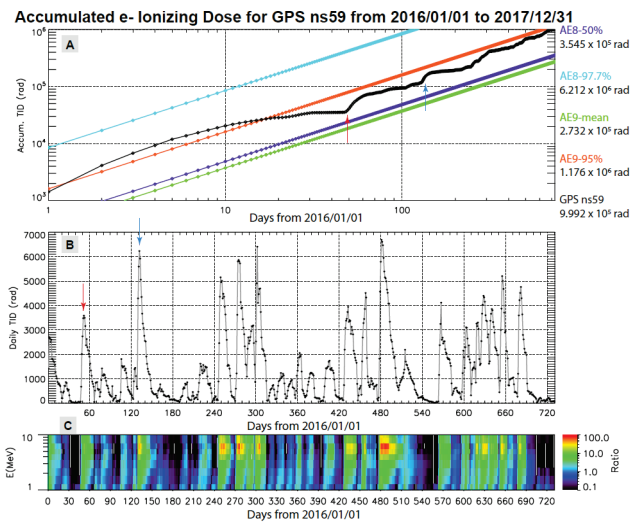

**Figure 4. Electron ionization doses accumulate over time for GPS ns59. A)** Accumulated doses calculated from measurements (black) and specification models. TID over the two years are given to the right side. **B)** Daily doses based on measurements. **C)** Measured electron fluxes normalized to mean values over two years.

Applying the same approach, we calculated ionization doses for protons. Measured proton fluence spectrum over 2016-2017 is plotted in Figure 5A, compared to other four spectra from trapped proton specification models AP8 and AP9 as well as the ESP model [7, 8] for transient solar protons: purple are calculated from AP8min fluxes (AP8min), aqua from AP8max (AP8max), green from AP9 mean values, and red from AP9 fluxes at 95% percentile. All specification model fluences are derived by providing GPS orbital parameters and temporal information to SPENVIS. Note the four model curves are overlapping to each other for energies > ~ 10 MeV due to the dominance of SEP particles determined by the ESP model. Fluence spectra in Figure 5A are used as inputs to GRAS for dose calculations.

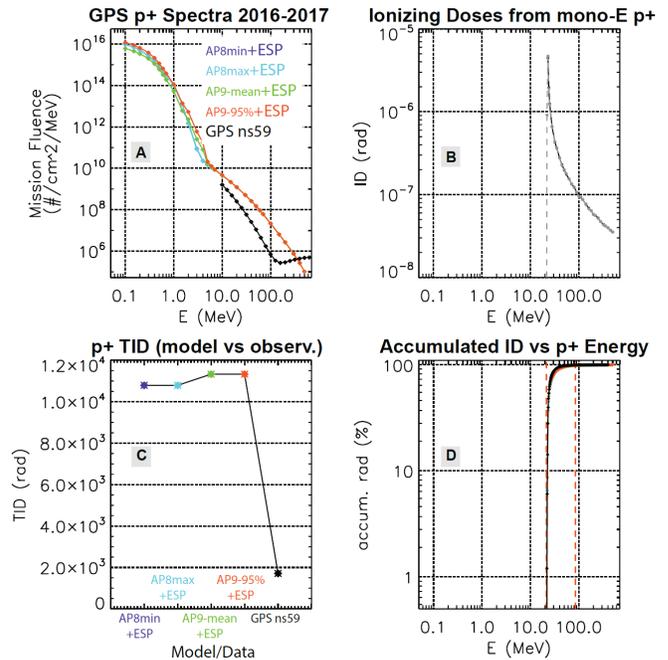

**Figure 5. Calculate proton TID over 2016-2017.** Panels are in the same format as Figure 3.

Figure 5B shows the proton dose contribution function curve for the specific geometry, calculated in the same way as electrons. Once more, the curve is like a reversed Bragg curve and can be understood from the shape of proton's LET curve inside Si: Protons below 20 MeV have ranges smaller than the 100 mil Al shielding and thus have no contribute to ionization dose, and the significant peak seen at ~ 21 MeV can be explained by the high LET of protons with residual energies of 100s keV, and then the steeply declining distribution extending beyond 100 MeV resemble the LET curve that decreases more than 20 times from 1 to 100 MeV.

Using the same superposition method, two-year proton TID values for ns59 are calculated and shown in Figure 5C. The proton TID from measured fluence is ~ 1.71 krad which is about 15% of the TID values (~ 11 krad) calculated from models. Here the dose values from protons are much lower than those from electrons. Doses from AP9 are slightly higher than those from AP8 values due to the extended energy coverage of AP9.

As in Figure 5D, proton effective energy range can be determined from the normalized radiation curves, and is ~ 21 – 85 MeV for protons in GPS orbit as marked out by the two red dashed vertical lines. Protons within this energy range are mainly SEP particles, which explains the similar TID values for all four models as shown in Panel C. In other words, Figure 5C basically compares TID from measurements and ESP model, and the contributions from trapped protons for this specific geometry can be ignored. It can be seen from Panel A that the measured fluence within this effective energy range is significantly lower than that from ESP model.

We also used the measurements by ns59 to calculate proton dDR and track how ionization dose being accumulated over time. Figure 6A plots the dose accumulation curve based on measurements in black, in which abrupt increments (two examples marked out by arrows) are seen to be caused by high dDRs (the two corresponding spikes) in Panel B and associated with proton events (the vertical strips) as in Panel C. Clearly, for the two high dDR examples, the second is due to the major SEP event on day 621 while the first does not have any corresponding SEP event and is likely caused by the contamination from an MeV electron event. In comparison, doses calculated from specification models accumulate with constant rates. Proton TID values from different inputs are presented in the right side of Panel A. The average proton dDR from measurements is 2.34 rad/day, compared to the dDR range of 14.8 – 15.5 rad/day from the four model combinations. The much lower proton TID from measurements are mainly caused by the unusually low solar activity levels during these two years, and we further explore this in the next Section.

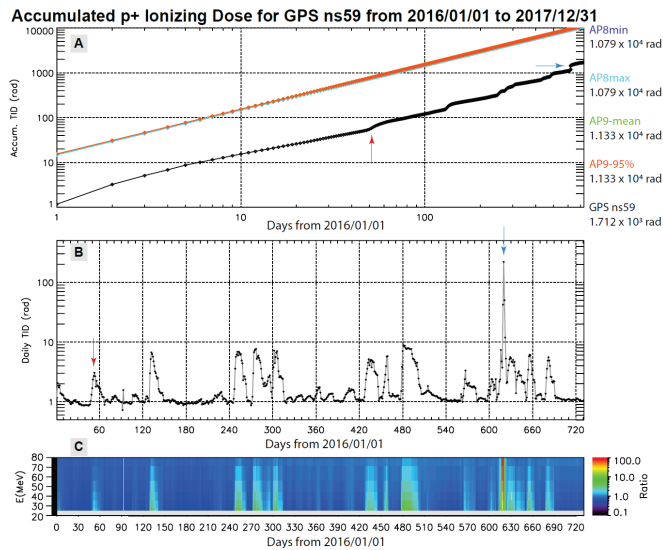

**Figure 6. Proton ionization doses accumulate over time for GPS ns59.** Same format as Figure 4.

## 4. DISCUSSIONS

Statistical studies on electron dDRs help us understand the distributions and cause of high daily doses. The occurrence distribution of dDRs in Figure 7A has two main components: One is an exponential distribution in the low dDR range of < ~ 1400 rad/day, which includes > 50% of the days; and the other is a wide distribution for high dDRs, in which the highest dDR is ~ 5 times of the mean dDR value. The relationship between dDRs and the Dst index is also depicted as in Panel B. There exists a general tendency that higher dDRs occur with more negative Dst values. This trend can be explained by the knowledge that most MeV electron events occur during the recovery phase of geomagnetic storms [19], meanwhile the large variations for individual days are also consistent with previous discoveries that storm intensities are not linearly related to the intensities of MeV electron events [20, 21]. A closer inspection on very high dDRs above 6 krad/day reveals they occur on the days in the recovery phase of moderate storms.

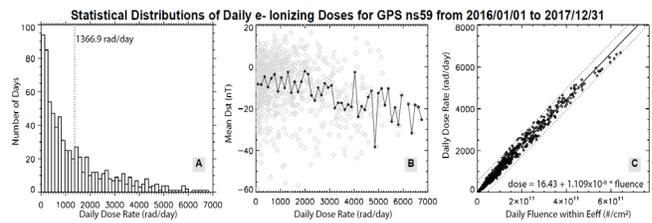

**Figure 7. Statistical characteristics of electron dDRs. A)** Occurrence distributions of dDRs. **B)** Relationship between dDRs and Dst index. The Black line is for averaged values and gray data points for individual days. **C)** Linear fitting between dDRs and electron effective fluence values, with the fitting function presented in bottom. The black fitting line locates between two gray dotted lines indicating error ranges of +/- 500 rad/day.

Moreover, a linear relationship can be identified between electron dDRs and daily fluences of electrons within the effective energy range (called effective fluence hereinafter), as shown in Figure 7C. The significance of this relationship is that it can conveniently be used for quick estimates of dose values directly from electron flux measurements, with no need of doing full scale radiation transport calculations.

Similar statistical studies were carried out for protons. It is clear from Figure 8A that low-level proton dDRs dominate the distribution. Data points in the first bin are mainly background, and data in the second bin are mostly from MeV electron contamination, as shown in Figure 6B. The limited data points in higher dose bins are real and from SEP events, in which the maximum dDR can be more than 100 times higher than quiet days. Proton dDRs have no significant relationship with geomagnetic indices including Dst and Kp since the trapped proton population barely contribute for the given geometry. Considering the wide range of proton dose values, a power-law relation between proton dDRs and the effective fluences of protons is identified as shown in Panel B. It should be noted that, since the fitting coefficient of 1.076 is very close to one, the relationship in Figure 8B can also be well represented by a linear function. Like electrons, this relationship can yield quick estimates of proton dose values directly from flux measurements.

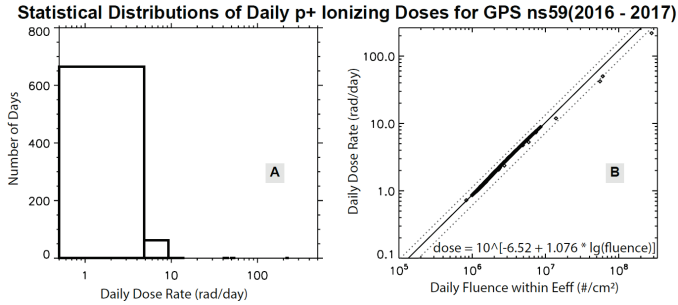

**Figure 8. Statistical characteristics of proton dDRs. A)** Occurrence distributions of dDRs. **B)** Fitting between dDRs and proton effective fluence values, with the fitting function presented in bottom. The black fitting line locates between two gray dotted lines indicating error ranges of +/- 30%.

GPS CXD data have been extensively inter-calibrated with measurements from other satellite missions [22, 23] and thus deemed reliable. However, the effects of non-isotropic directional distributions [24] and limited instrumental field-of-view (FOV) need to be considered. The CXD is mounted on the nadir panel of each three-axis stabilized GPS satellite, with instrumental axis always pointing towards the Earth. This looking direction strays farther away from the local 90° pitch angle when satellites moving to higher latitudes. In addition, distributions of the collimating holes on CXD particle sensors define limited FOV [16]—about 15% of the full $4\pi$ solid angle—and thus only cover a small portion of the local pitch angles. Therefore, we need to estimate what the difference can be between the electron intensities measured by CXD and the omni-directional electrons experienced by GPS ns59 satellite. To address this question, we compared CXD data to measurements from one RBSP satellite [25] in 2014, during which a list of MeV electron events and SEP events occurred, by taking advantage of the full pitch-angle coverage in RBSP particle measurements.

Electrons are first compared as in Figure 9. The top panel shows daily averaged and directionally averaged fluxes (meaning omni-directional fluxes divided by the full $4\pi$ solid angle) of 2 MeV electrons in-situ measured by Relativistic Electron-Proton Telescope (REPT) [26] on board RBSP-A. REPT measures electrons ranging from ~2-20 MeV. There are quite a few intense MeV electron events (yellow and red regions) observed in 2014, particularly during the fourth quarter. Using the level 3 pitch-angle resolved data, REPT fluxes can be mapped along magnetic field lines and projected from near-equatorial RBSP locations up to the GPS ns59 orbit positioned at high latitudes for local pitch-angle distributions. In this way, local directionally averaged fluxes at GPS locations can be derived and shown in Panel B. Differences can be seen at high L-shells between fluxes in Panels A and B due to non-isotropic pitch-angle distributions. These projected fluxes are further compared to those in-situ measured by CXD (shown in Panel C), with the flux ratios presented in Panel D. Most of the time the ratios are close to the unit in blue color, but dynamics are also visible associated with geomagnetic activities (Panel E), particularly during intense MeV electron events at high L-shells. To get a general idea of flux ratio distributions, we plot the mean ratios over 2014 as a function of L-shell and energy in Panel F. Clearly, the ratios grow with the L-shell as well as energy, with the highest value (22.7) in the upper right corner of the panel and the lowest value of 1.36 in lower left. Since the overlapping L-shells are limited between GPS and RBSP, we did not intend to get an accurate ratio factor between ionization doses, which requires to consider both the whole electron spectrum and extrapolation to higher L-shells. Instead, further averaging the flux ratio values in Panel F within the energy range of [2, 4] MeV, we have a mean ratio value of ~4.1, which gives us an estimate of how much CXD fluxes may underestimate the "real" electron TID.

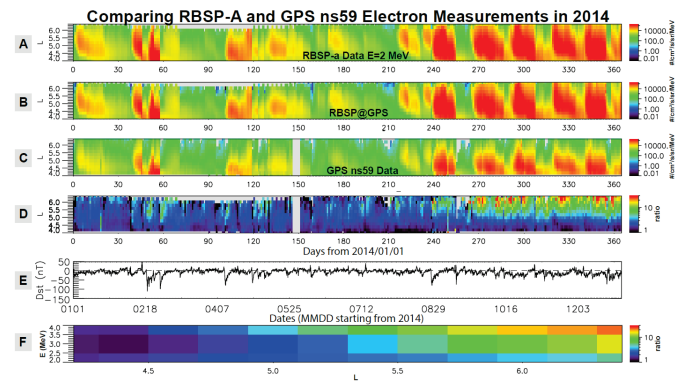

**Figure 9. Compare GPS ns59 and RBSP-A electron fluxes in 2014. A)** Daily averaged and directionally averaged fluxes of 2 MeV electrons in-situ measured by RBSP-A are plotted as a function of L-shell and time. **B)** Local directionally averaged fluxes projected from RBSP-A orbit to ns59 locations. **C)** 2 MeV electron fluxes in-situ measured by CXD on board ns59. **D)** Ratios between fluxes in Panels B and C. **E)** Dst index. **F)** Averaged ratios between projected fluxes and in-situ CXD measurements are plotted as a function of energy and L-shell.

Although protons contribute much less than electrons to TID, it is interesting to compare protons in the similar way and results are shown in Figure 10. Panel A presents protons measured by Relativistic Proton Spectrometer (RPS, [27]) instrument carried by the RBSP-A. The lack of radial features in the distributions in Panel A indicates the dominance of instrument background, except for the six SEP intervals registered with high flux values. Effects from MeV electrons are also discernible in RPS data, particularly in the last quarter of the year. Then, directionally resolved

RPS fluxes are magnetically projected to GPS ns59 locations, as shown in Panel B. Indeed, due to the almost isotropic directional distributions in the background, fluxes in Panel B barely show any difference compared to those in Panel A. Panel C plots CXD in-situ measured fluxes. Ratios between projected fluxes and in-situ measured fluxes are shown in Panel D, in which large ratio values (>30) dominate, indicating the large difference between two instrument backgrounds. For the major SEP event b, flux ratios for L-shells between 4.6 (the max penetration at GPS) and 6.2 have a mean value of 2.4 for 60 MeV protons. This suggests nearly isotropic directional distributions in solar protons during major SEP event and is consistent with the comparison in another major SEP event in 2017 [14]. However, for the weaker SEP events, large flux ratios reach up to ~10, which indicates significant directional differences. Therefore, ratio factors of ~2 - 10 are what needs to be considered when one calculates solar proton doses using GPS measurements.

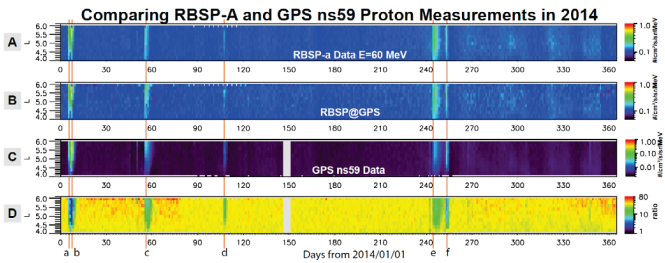

**Figure 10. Compare GPS ns59 and RBSP-A proton measurements in 2014.** **A)** Daily averaged and directionally averaged fluxes of 60 MeV protons in-situ measured by RBSP-A are plotted as a function of L-shell and time. **B)** Local directionally averaged fluxes projected from RBSP-A orbit to ns59 locations. **C)** 60 MeV protons in-situ measured by CXD on board ns59. **D)** Ratios between fluxes in Panels B and C.

Based on the long-term GPS electron data, we were able to quickly calculate yearly dose rates (yDRs) and study their trend over the full solar cycle 24 (starting from the end of 2008). Figure 11 presents the yDRs (placed on the middle of each corresponding year) and variations experienced by ns59 from 2006 to 2018. It is clear that 2015, 2016 and 2017 are the peak years for yDRs for this solar cycle (the second peak is 2012), which is consistent with the knowledge that MeV electron levels are generally the highest during the declining phase of solar cycles [28]. More importantly, the average yDR based on measurements over solar cycle 24 is 237. krad/yr, which is about half of the average yDR of 500. krad/yr over 2016-2017 from measurements, and is 1.3 (1.7) times of the 177. (136.) krad/yr from AE8 (AE9) model average fluxes. These high yDRs from measurements were not expected since the solar cycle 24 is known to be unusually quiet. Obviously, the doses calculated based on GPS measurements add plentiful details that the statistical models lack.

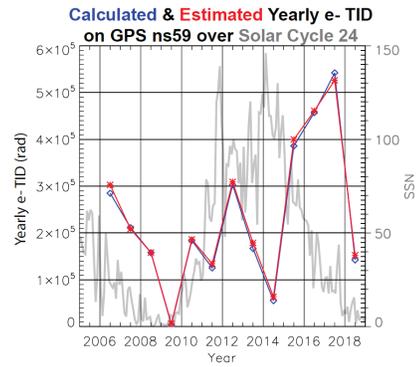

**Figure 11. Electron yearly ionization doses experienced by GPS ns59 over thirteen years (2006-2018).** Doses for data points in blue are calculated from in-situ measurements, and doses in red are estimated from effective fluences using the statistical relation determined in Figure 7C. Sunspot numbers are plotted in gray.

Moreover, it is interesting to compare the ionizing doses on GPS to those on GEO as in [3] since the same geometry being used. For electron TIDs, GPS ns59 has an averaged yDR of 237. krad/year while LANL-04A in GEO has a much lower value of 17.3 krad/year. Although it seems natural since GPS travel through the heart of outer electron belt, it is worth pointing out that measurement-based TID on GPS is more than 2 times of those from AE8/AE9 model median/mean fluences, while measurement-based TID on GEO is about 80% or even lower than those from AE8/AE9 models. This significant difference suggests that the margin factors on empirical models can vary largely over different orbits. As for protons, the highest dDR is ~220 rad/day for GPS during the major SEP in 2017, while GEO has a maximum dDR of > 10 krad/day in the same SEP event. This significant difference in proton TID can be explained by the better protection from geomagnetic field on the GPS orbit than GEO.

Results from this study provide us a hint of the margin factors for other MEOs. Indeed, along with the flux ratio factors shown in Figure 9, the publically available GPS particle measurements can be potentially used on other MEOs to estimate ionizing doses. We leave this to our future work.

Finally, we have used a much simplified solid sphere geometry for this demonstrative study. Indeed, the high yDR values derived here strongly suggest the necessity of thicker shielding so as to reduce the TID in the GPS orbit. Nevertheless, the method developed in this work can be easily extended to any realistic geometry with different shielding designs for satellites in MEO and other orbits. Dose calculations of this kind will not only provide out-of-sample validation for space radiation specification models, but also deliver detailed DRs in a timely manner that are needed by satellite operation and space system diagnosis.

## 5. SUMMARY AND CONCLUSIONS

Long-term GPS particle datasets have been widely used in numerous studies leading to many significant scientific discoveries and space applications, including understanding the dynamics of relativistic electrons and analyzing radiation effects (e.g., on the internal charging [29]). In this study, we show case that this valuable data set can also be extended for total ionization dose and dose rate calculations that verify and supplement those calculated from specification models.

Specifically, this study has used GRAS model and the in-situ electron and proton measurements from GPS ns59 satellite for determining TIDs and DRs. For the specific solid sphere geometry, a superposition method has been developed to speed up ionization dose calculations. After obtaining dose contribution functions, it was shown that ionization doses are closely related to particle fluences within the effective energy range of ~ 1.3 – 4 MeV for electrons and 21 – 85 MeV for protons in GPS orbit. For electrons over the solar cycle 24, yDRs from measurements range from 4.05 (in 2009) to 542. (2017) krad/yr with a mean value of 237. krad/yr. This mean is about ~1.5 times of the yDR values from AE8 and AE9 models. Particularly for the two peak years 2016-2017, mean electron yDR is 500. krad/yr, compared to the mean yDR of 177. and 136. krad/yr from AE8 and AE9 median/mean distribution; and average electron dDR from measurements is 1.37 krad/day, compared to the mean dDR of 0.485 and 0.374 krad/day from AE8 and AE9 mean fluxes. The highest dDR from measurements during 2016-2017 is above 6 krad/day, which is within the range of 8.50 krad/day rate set by AE8-97.7% but well beyond the dDR of 1.61 krad/day set by AE9-95%. Comparisons to RBSP electron data suggest that ionization doses from electrons in MEO can be ~4 times higher than those calculated from GPS measurements. For protons during the same two-year interval, the TID from measured fluence is ~ 1.71 krad which is about 15% of the TID values (~ 11 krad) calculated from models. Averaged proton dDR from measurements is 2.34 rad/day, also much lower than the dDR range of 14.8 – 15.5 rad/day from the four model combinations. Proton contributions to TID in MEO are dominated by solar protons and play a minor role compared to electrons.

## ACKNOWLEDGEMENTS

This work was performed under the auspices of the U.S. Department of Energy and also supported by the Laboratory Directed Research and Development (LDRD) program at LANL (award 20190262ER). We are grateful for the use of GRAS code developed by European Space Agency.

## Biography

**Dr. Yue Chen** received his PhD degree in space physics from Rice University, Houston, TX in 2003. He has been with LANL for more than 15 years, leading and participating in multiple projects aiming to understand, model, and predict the dynamics of Van Allen radiation belts. He also works on quantifying space radiation effects on satellites.

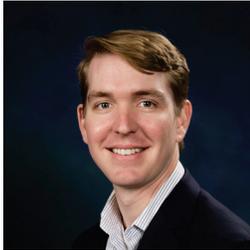

**Dr. Matt Carver** is a Lead Scientist at Booz Allen Hamilton working on their National Agencies account. He received a Ph.D. in physics from the University of Florida in 2016 conducting high energy physics research on the Compact Muon Solenoid experiment searching for evidence of new subatomic particles. At Booz Allen he applies data science and machine learning concepts to internal R&D efforts for anomaly detection, collision avoidance, trajectory propagation, and pattern recognition. Prior to this Matt was at LANL for 4 years where he focused on analyzing data from energetic charged particle and x-ray sensors onboard GPS satellites as well as development of scientific modeling code to perform system level simulations of the United States Nuclear Detonation Detection System.

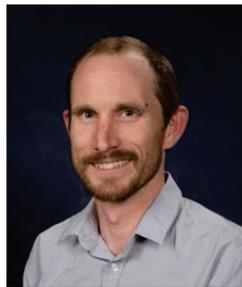

**Dr. Steven K. Morley** is a scientist in the Space Science and Applications group at Los Alamos National Laboratory (LANL), where he has worked for over 10 years. Prior to joining LANL he worked at the University of Newcastle, Rutherford Appleton Laboratory, and British Antarctic Survey. He received his M. Phys in Physics with Astronomy (2000) and Ph.D. in Physics (2004) from the University of Southampton. Their work is primarily focused on physics of the solar wind-magnetosphere-ionosphere system, and its application in the space weather domain. They have contributed to scientific satellite missions including the microsatellite FedSat and NASA's Magnetospheric Multiscale. His research interests include electron radiation belt dynamics, solar energetic particle events, magnetospheric substorms, data-model fusion, and numerical space weather prediction.

**Dr. Andrew Hoover** is a member of the scientific staff at Los Alamos National Laboratory and received his Ph.D. in Physics at New Mexico State University in 2003, with a focus on experimental high-energy physics. He is presently the Chief Scientist for the combined x-ray dosimeter (CXD) instrument, a component of the U.S. nuclear detonation detection system (USNDS). Prior research includes superconducting ultra-high resolution x-ray and gamma-ray sensors, gamma-ray imaging, gamma-ray bursts, and a wide variety of simulation and modeling efforts for hard radiation instruments.

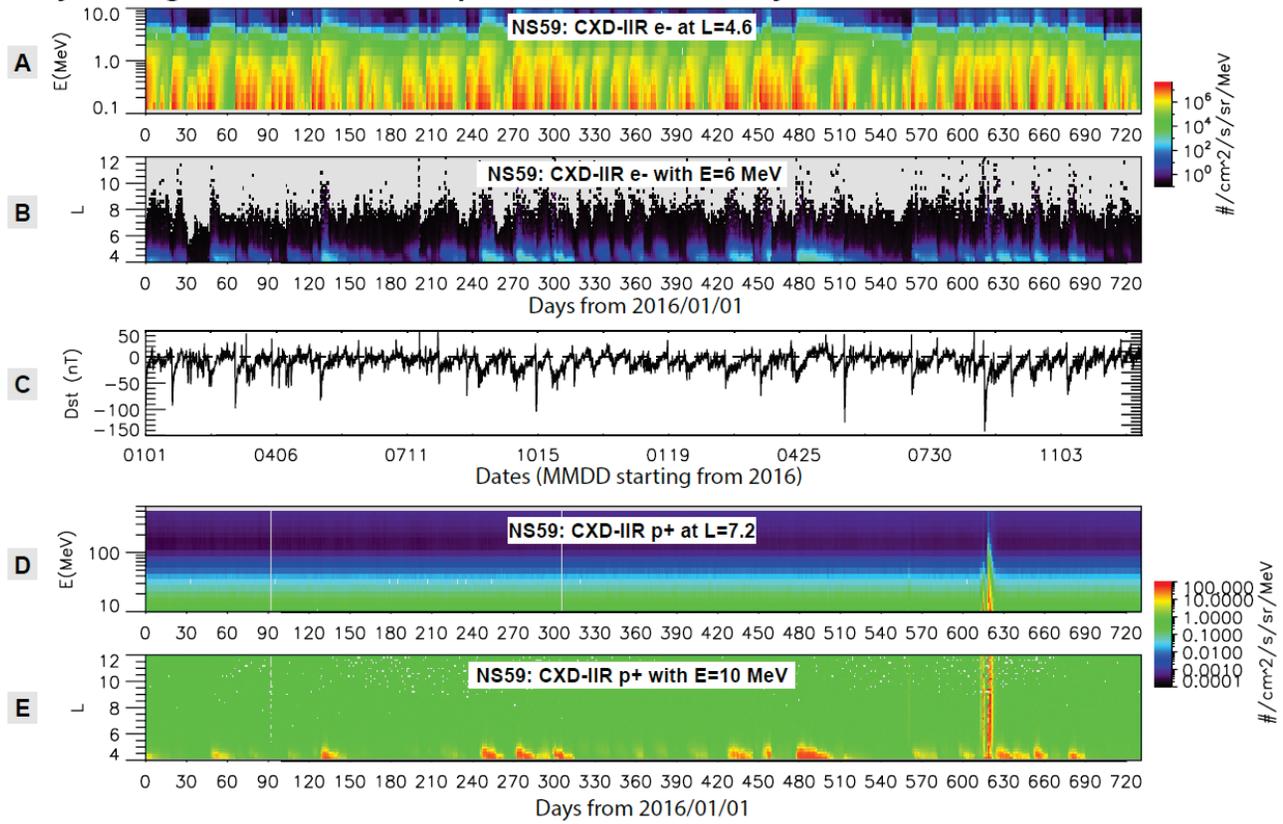

**Figure 1. Overview of electron and proton dynamics observed by GPS ns59 in 2016 and 2017.** Panels A and B (D and E) are electron (proton) fluxes sorted as a function of energy and L-shell, and Panel C shows the Dst index. Data gaps exist for both electrons and protons (e.g., on day 91).

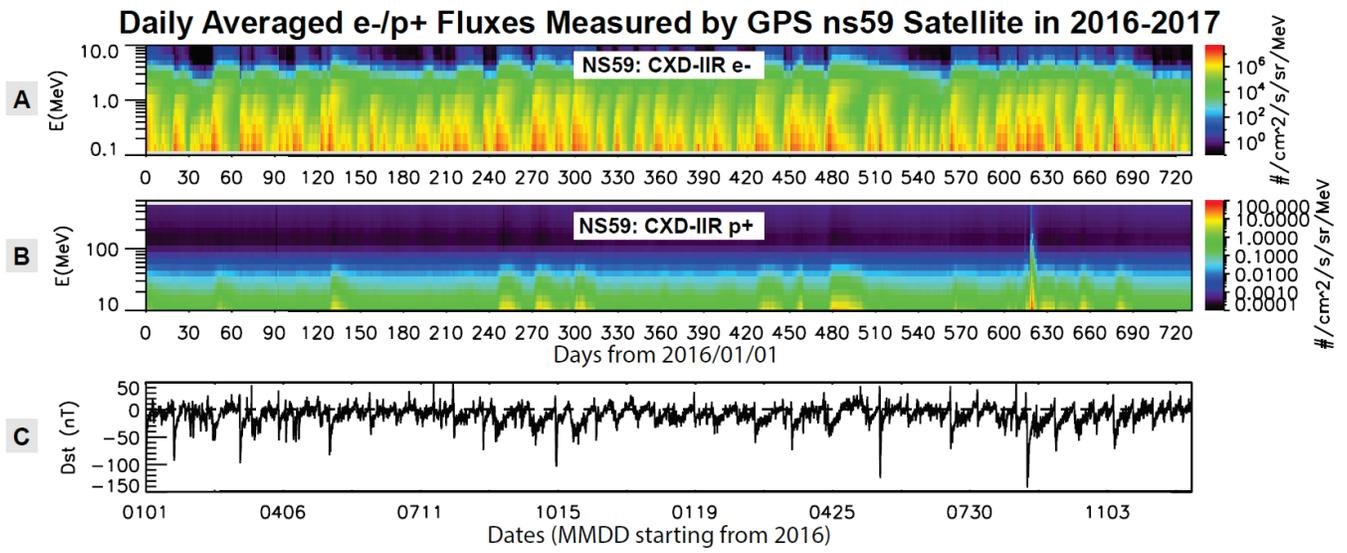

**Figure 2. Overview of L-shell and daily averaged electron and proton fluxes observed by GPS ns59 in 2016 and 2017.** Panel A (B) are electron (proton) daily fluxes averaged over L-shells, and Panel C shows the Dst index.

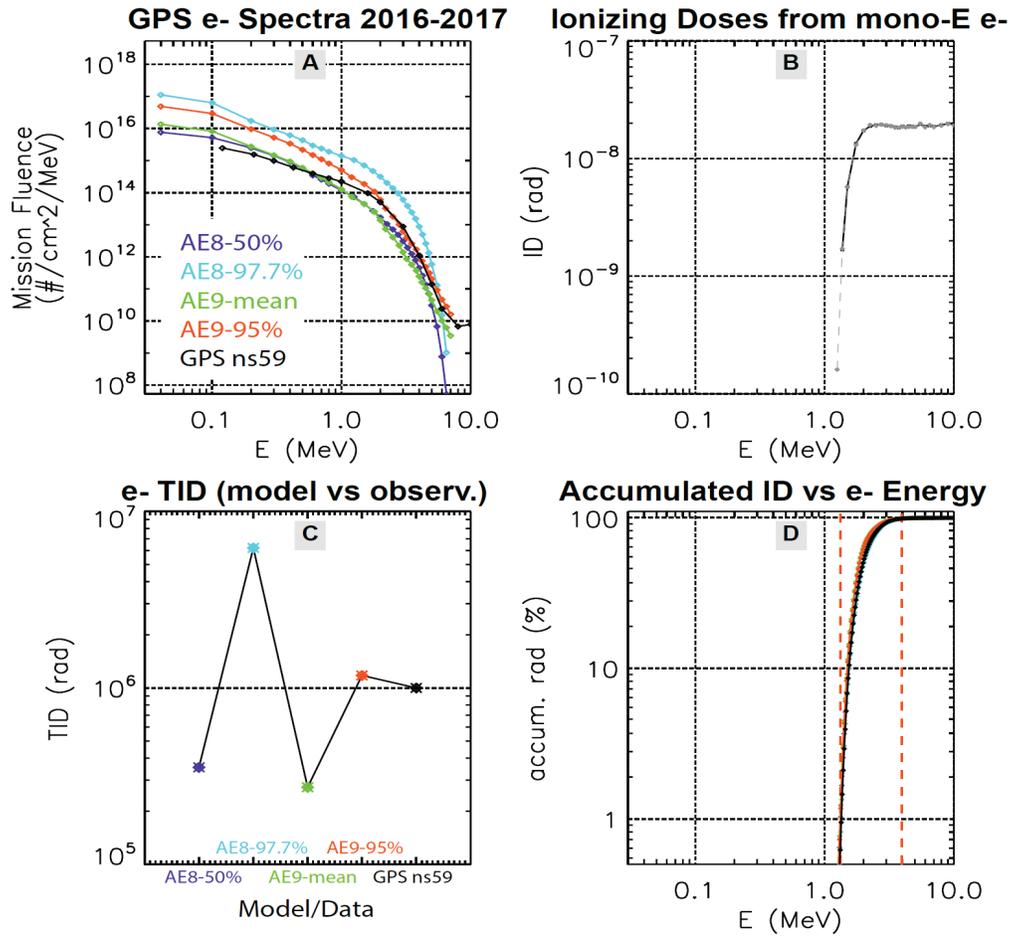

**Figure 3. Calculate electron TID over 2016-2017. A)** Electron fluence spectra, **B)** dose contribution function curve, **C)** TID calculated from different electron fluence inputs, and **D)** effective energy range determined from normalized accumulative dose percentage curves.

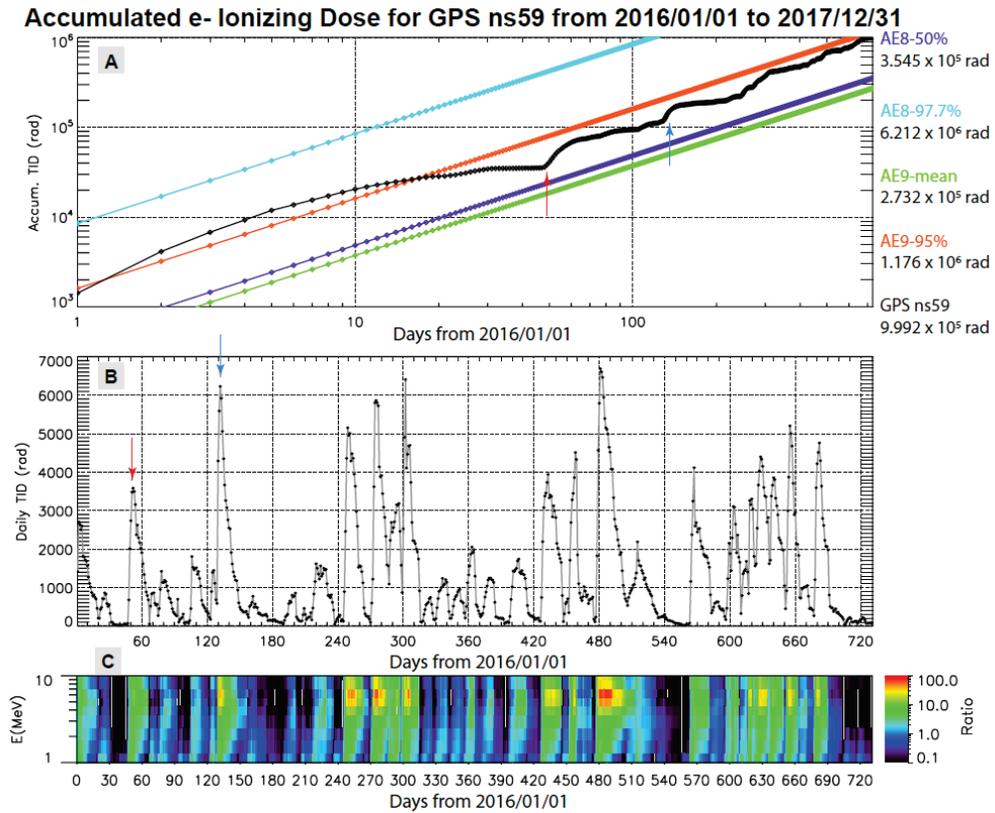

**Figure 4. Electron ionization doses accumulate over time for GPS ns59. A)** Accumulated doses calculated from measurements (black) and specification models. TID over the two years are given to the right side. **B)** Daily doses based on measurements. **C)** Measured electron fluxes normalized to mean values over two years.

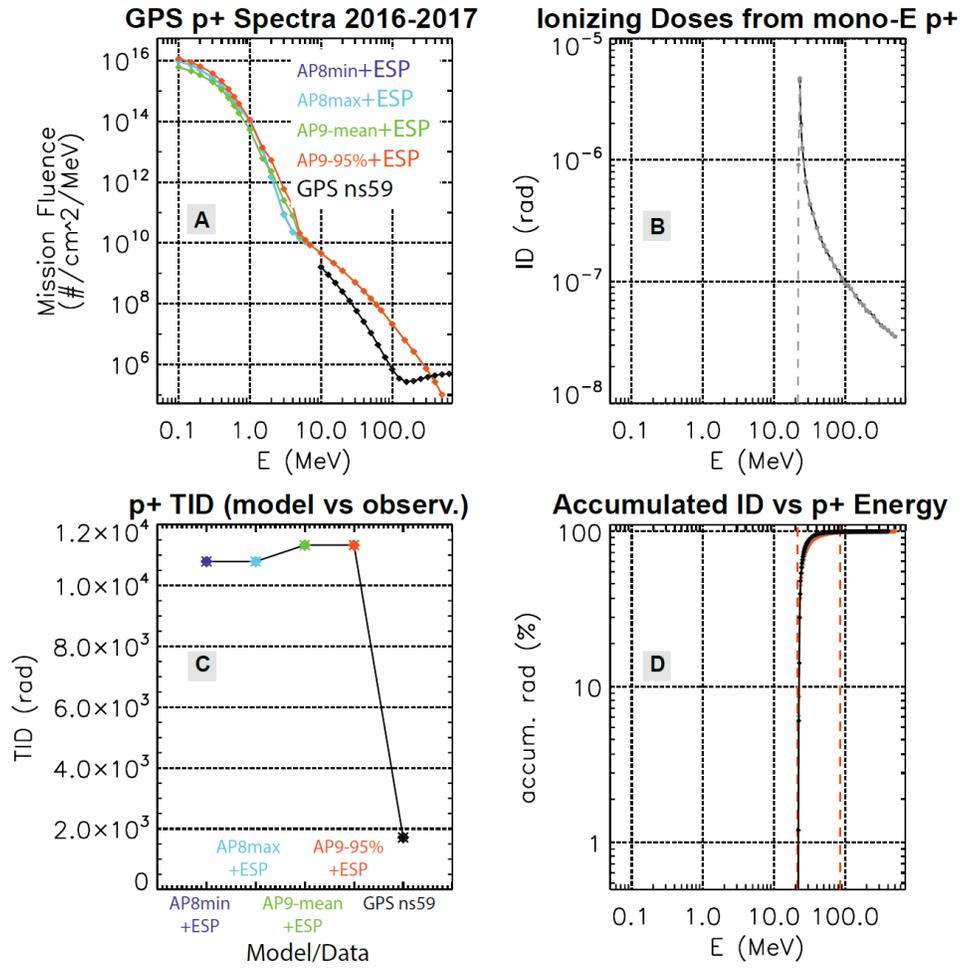

**Figure 5. Calculate proton TID over 2016-2017.** Panels are in the same format as Figure 3.

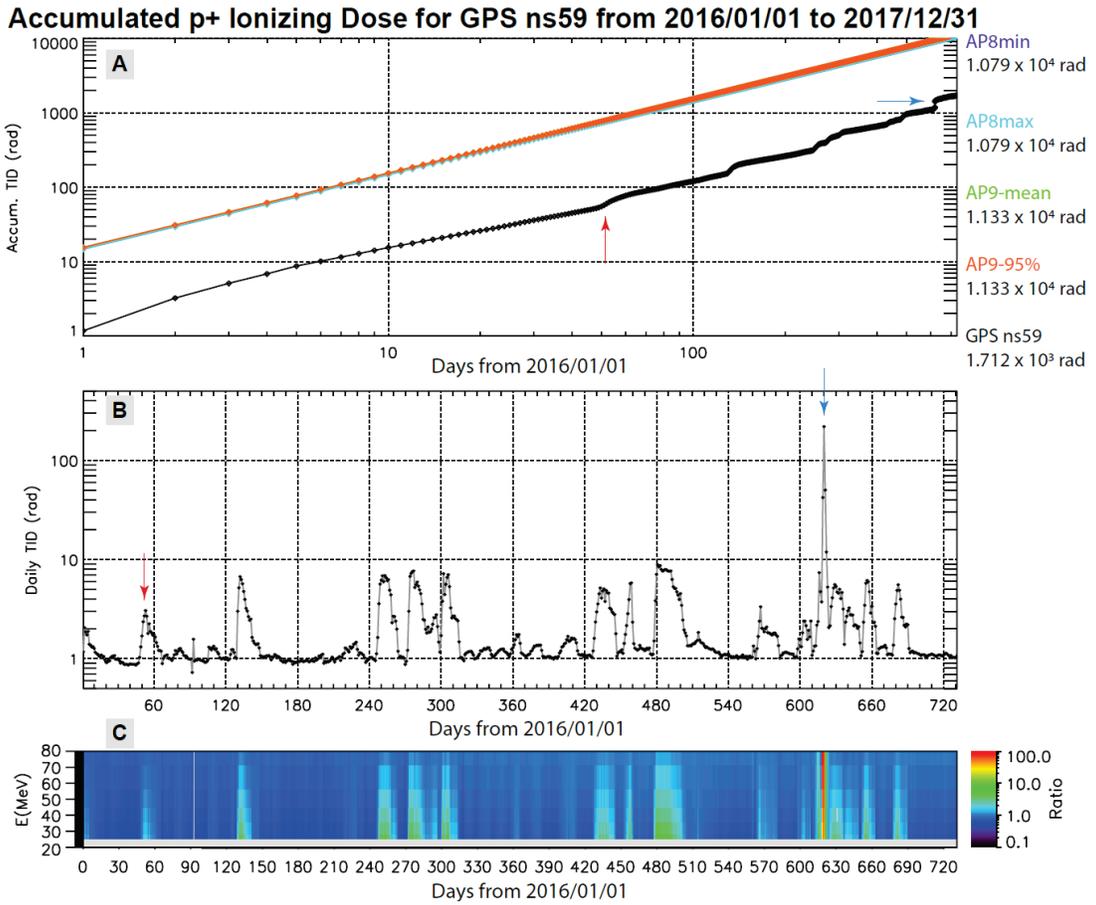

**Figure 6. Proton ionization doses accumulate over time for GPS ns59.** Same format as Figure 4.

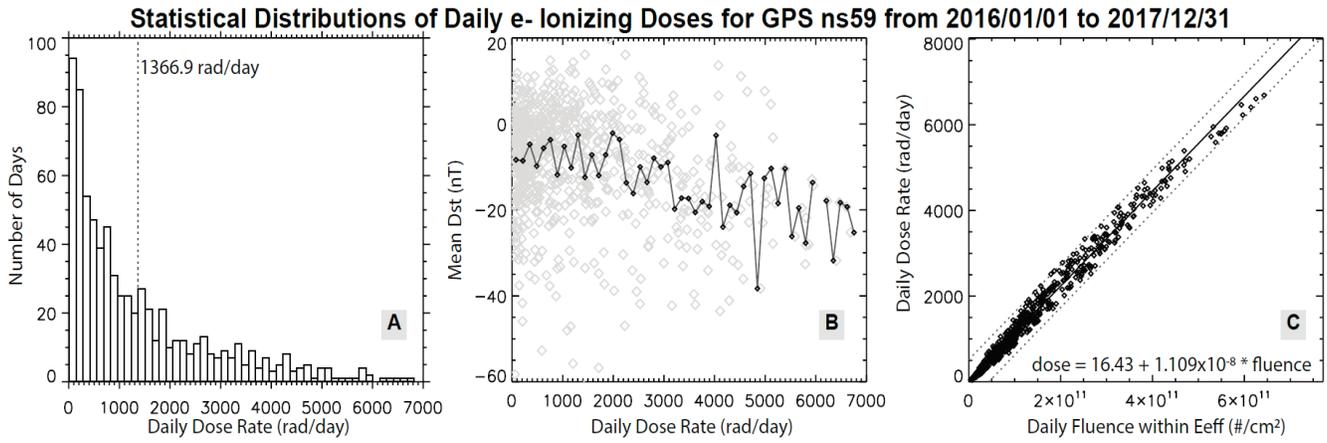

**Figure 7. Statistical characteristics of electron dDRs. A)** Occurrence distributions of dDRs. **B)** Relationship between dDRs and Dst index. The Black line is for averaged values and gray data points for individual days. **C)** Linear fitting between dDRs and electron effective fluence values, with the fitting function presented in bottom. The black fitting line locates between two gray dotted lines indicating error ranges of +/- 500 rad/day.

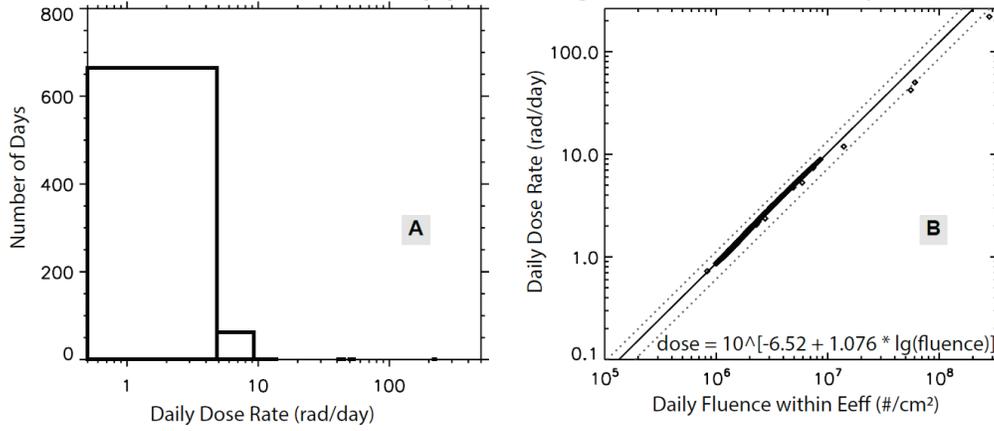

**Figure 8. Statistical characteristics of proton dDRs. A)** Occurrence distributions of dDRs. **B)** Fitting between dDRs and proton effective fluence values, with the fitting function presented in bottom. The fitting line is in black located between two gray lines indicating error ranges of +/- 30%.

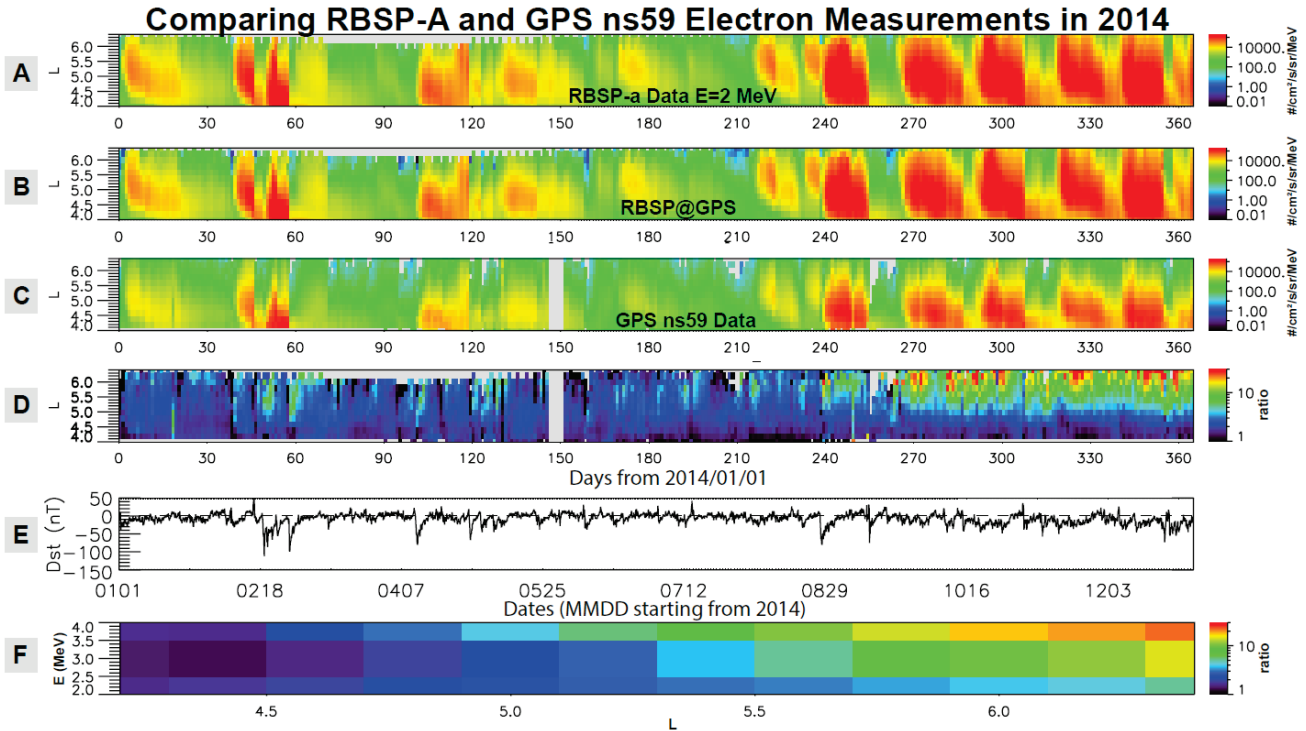

**Figure 9. Compare GPS ns59 and RBSP-A electron fluxes in 2014. A)** Daily averaged and directionally averaged fluxes of 2 MeV electrons in-situ measured by RBSP-A are plotted as a function of L-shell and time. **B)** Local directionally averaged fluxes projected from RBSP-A orbit to ns59 locations. **C)** 2 MeV electron fluxes in-situ measured by CXD on board ns59. **D)** Ratios between fluxes in Panels B and C. **E)** Dst index. **F)** Averaged ratios between projected fluxes and in-situ CXD measurements are plotted as a function of energy and L-shell.

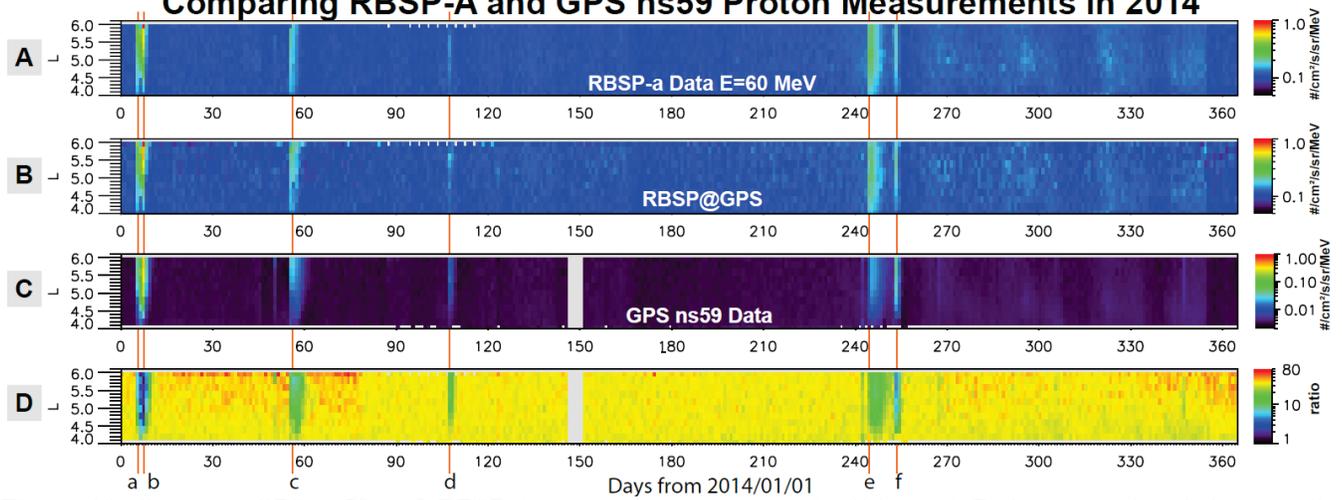

**Figure 10. Compare GPS ns59 and RBSP-A proton measurements in 2014.** **A)** Daily averaged and directionally averaged fluxes of 60 MeV protons in-situ measured by RBSP-A are plotted as a function of L-shell and time. **B)** Local directionally averaged fluxes projected from RBSP-A orbit to ns59 locations. **C)** 60 MeV protons in-situ measured by CXD on board ns59. **D)** Ratios between fluxes in Panels B and C.

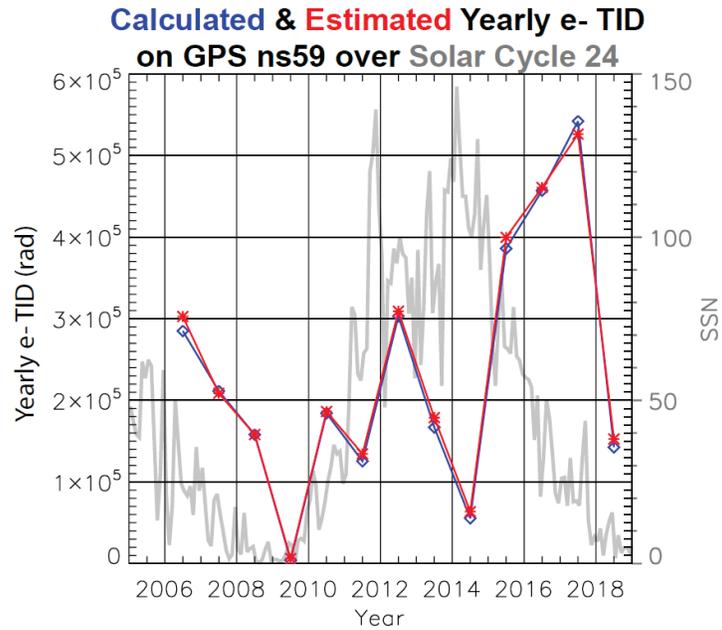

**Figure 11. Electron yearly ionization doses experienced by GPS ns59 over thirteen years (2006-2018).** Doses for data points in blue are calculated from in-situ measurements, and doses in red are estimated from effective fluences using the statistical relation determined in Figure 7C. Sunspot numbers are plotted in gray.